\documentclass[pra,twocolumn,epsfig,rotate,superscriptaddress,showpacs]{revtex4}
\usepackage[dvips]{graphics}
\usepackage{graphicx}
\usepackage{amsfonts}
\usepackage{amssymb}
\usepackage{amsmath}
\usepackage{subfigure}
\usepackage{prettyref}
\usepackage{float}
\usepackage{amsmath}
\usepackage{amssymb}
\usepackage{esint}

\begin{document}


\title{Protecting unknown two-qubit entangled states by nesting Uhrig's dynamical decoupling sequences}
\author{Musawwadah Mukhtar}
\affiliation{Department of Physics, National University of Singapore, 117542, Republic of Singapore}
\author{Wee Tee Soh}
\affiliation{Department of Physics, National University of Singapore, 117542, Republic of Singapore}
\author{Thuan Beng Saw}
\affiliation{Department of Physics, National University of Singapore, 117542, Republic of Singapore}

\author{Jiangbin Gong}
\email[]{phygj@nus.edu.sg}
\affiliation{Department of Physics, National University of Singapore, 117542, Republic of Singapore}
\affiliation{Centre for Computational Science and Engineering,
\\
National University of Singapore, 117542, Republic of Singapore}
\affiliation{NUS Graduate School for Integrative Sciences and Engineering, Singapore
117597, Republic of Singapore}
\date{\today}

\begin{abstract}
Future quantum technologies rely heavily on good protection of quantum entanglement against environment-induced decoherence.  A recent study showed that an extension of Uhrig's dynamical decoupling (UDD) sequence can (in theory) lock an arbitrary but known two-qubit entangled state to the $N$th order using a sequence of $N$ control pulses [Mukhtar {\it et al.}, Phys.~Rev.~A {\bf 81}, 012331 (2010)]. By nesting three layers of explicitly constructed UDD sequences, here we first consider the protection of unknown two-qubit states as superposition of two known basis states, without making assumptions of the system-environment coupling. It is found that the obtained decoherence suppression can be highly sensitive to the ordering of the three UDD layers and can be remarkably effective with the correct ordering. The detailed theoretical results are useful for general understanding of the nature of controlled quantum dynamics under nested UDD. As an extension of our three-layer UDD, it is finally pointed out that a completely unknown two-qubit state can be protected by nesting four layers of UDD sequences.
  This work indicates that when UDD is applicable (e.g., when environment has a sharp frequency cut-off and when control pulses can be taken as instantaneous pulses), dynamical decoupling using nested UDD sequences is a powerful approach for entanglement protection.
\end{abstract}

\maketitle

\section{Introduction}
Virtually all quantum systems are coupled to an environment and hence suffer from decoherence.  Even more troublesome, the implication of decoherence for entangled states is far more severe than for a single quantum system. For example,  even though a superposition state of one qubit cannot be completely decohered within any finite time, the entanglement between two such qubits may be totally destroyed by decoherence within a very short time~\cite{karol,science,deathexp}.  Evidently then, developing useful schemes to protect quantum entanglement from environment-induced decoherence is crucial for entanglement-based quantum technologies.

One promising approach towards decoherence suppression is dynamical
decoupling (DD)~\cite{DD}, which advocates the application of a
sequence of instantaneous control pulses to effectively average out
the system-environment coupling.  Important extensions of the
original DD approach have also been developed, e.g., ``concatenated
dynamical decoupling" pulses~\cite{cDD}, soft but optimized
pulses under an energy cost constraint~\cite{kurizki} or a minimal
leakage requirement~\cite{LAwu}, and Uhrig's dynamical decoupling
(UDD)~\cite{UDD1,UDD2,UDD3} sequence that can achieve a very high
efficiency of decoherence suppression, i.e., decoherence suppression
to the $N$th order with a sequence of $N$ instantaneous control pulses.
Although UDD requires a sharp frequency cut-off in the bath spectrum
\cite{davidson,newexp,sarma} and cannot operate well if we set
limitations on the control pulse width \cite{viola10}, UDD has
attracted substantial interests soon after its discovery.  In addition to
its high-order suppression of decoherence (at least in theory), UDD for single-qubit decoherence
suppression could be powerful because it works for most general
system-bath coupling \cite{UDD3}, for a bath that has unknown
spectral density (but with a sharp cut-off), and for time-dependent
system-bath Hamiltonians as well \cite{uhrigJA}.  Experimental
studies of UDD under two specific situations have been reported ~\cite{exp1,exp2,exp3}.  For
a recent concise review on UDD-related theoretical studies, see Ref.
\cite{Liureview}.

So far the majority of DD studies have focused on single-qubit systems \cite{liunewwork}.
Hence it is urgent to investigate if quantum entanglement, e.g.,
 two-qubit entangled states, can be well protected by DD.  Because preserving two-qubit
 entanglement is more subtle than preserving single-qubit coherence, we wish to find a general
 control scheme to achieve good entanglement preservation without assuming any particular form of
 system-environment coupling. Indeed, given so many different ways of coupling a composite quantum system
 to an environment, in general a specific assumption about system-environment coupling may  over-simplify the
 issue of entanglement protection.  As such,  a universal and efficient DD scheme for entanglement
 protection should be of sufficient interest.
 Certainly, if under certain environments some crucial information about system-environment coupling
 becomes available, then a general DD scheme may be further reduced, a situation exploited in
  the first experimental study of entanglement protection by DD in a solid-state environment~\cite{wangya}.

In our early study~\cite{pra1}, it was shown that an extended UDD can also lock a known two-qubit
 entangled state to the $N$th order with $N$ control pulses.
The present study is concerned with the protection of unknown entangled states, a situation
that is more relevant for quantum information processing. To that end we extend the recent work
by West {\it et al.}~\cite{west}, where a scheme based on two layers of UDD sequences is
proposed to suppress both the population relaxation and transverse dephasing of one single qubit.
 In particular, we show that by nesting three layers of UDD sequences, it is possible to lock a two-qubit
 entangled state as an unknown superposition of two basis states, to the $N$th order, using about $N^3$ control
 pulses in total. This entanglement protection scheme is independent of how the two-qubit system
 is coupled to its environment. The control operators used in our nested UDD are also explicitly
 constructed based on two arbitrary, but known, basis states.  We shall also show that there are different
 scenarios in constructing the control operators.  Interestingly, it is found that the
 ordering of the nested UDD layers can be a crucial factor for achieving high-order entanglement protection.  This intriguing ordering dependence is absent in two-layer UDD for single-qubit decoherence control, thus offering more insights into quantum decoherence control via nested UDD.

 As a further extension of our three-layer UDD, we show that it is possible
 to construct a four-layer UDD scheme such that a totally unknown
 two-qubit state can be protected using about $N^4$ pulses of single-qubit control operators.  Experimentally, this might be even more challenging than
realizing $N^3$ pulses of two-qubit operators. However, such a theoretical possibility is hoped to motivate future studies.

This paper is organized as follows. In Sec. II, after introducing
the most general system-environment coupling for two-qubit systems, we briefly review our previous
extension of UDD from one-qubit to  two-qubit systems. Emphasis is
placed on the key requirements to achieve such an extension.   In
Sec. III, we consider two schemes for nesting UDD in three layers in order to protect unknown
two-qubit entangled states. Supporting numerical results are also
presented. Section IV discusses a four-layer UDD scheme, followed by Sec. V that concludes this paper.

\section{Protecting a known two-qubit state by UDD}
\subsection{General total Hamiltonian of a two-qubit system interacting with a bath}
In terms of system-environment coupling, two-qubit systems are far more complex than one-qubit systems.
A general total Hamiltonian describing a two-qubit system interacting with a bath can be written as
\begin{eqnarray}
H
 & = & c_0+\sigma_{x}^{1}c_{x,1}+\sigma_{y}^{1}c_{y,1}+\sigma_{z}^{1}c_{z,1}
 + \sigma_{x}^{2}c_{x,2} \nonumber \\
 && +\ \sigma_{y}^{2}c_{y,2}+\sigma_{z}^{2}c_{z,2} +\sigma_{x}^{1}\sigma_{x}^{2}c_{xx}+\sigma_{x}^{1}\sigma_{y}^{2}c_{xy}\nonumber \\
& & +\ \sigma_{x}^{1}\sigma_{z}^{2}c_{xz}+\sigma_{y}^{1}\sigma_{x}^{2}c_{yx}+\sigma_{y}^{1}\sigma_{y}^{2}c_{yy}
 +\sigma_{y}^{1}\sigma_{z}^{2}c_{yz}\nonumber\\
 && +\ \sigma_{z}^{1}\sigma_{x}^{2}c_{zx}+\sigma_{z}^{1}\sigma_{y}^{2}c_{zy}
 +\sigma_{z}^{1}\sigma_{z}^{2}c_{zz}.
\label{sb-H}\end{eqnarray} For convenience each term in the above
total Hamiltonian is assumed to be time independent (this assumption
can be lifted). Here $c_0$ represents the self-Hamiltonian of
the bath, $\sigma_x^{j}$, $\sigma_{y}^{j}$, and $\sigma_z^{j}$ are
the standard Pauli matrices for the first ($j=1$) or the second
($j=2$) qubit, $c_{\gamma,j}$ and $c_{\gamma\delta}$ ($\gamma,\delta=x,y,z$) represent
arbitrary smooth bath operators.  From Eq.~(1), it
is seen that in general, the bath may interact with each individual
qubit, or modulate the mutual interaction between the two qubits.
The latter situation naturally arises if, for example, the bath can
induce phonon excitations in a solid and hence perturb the
relative distance between the two qubits embeded in the solid. The
above total Hamiltonian is in the most general form, because it can
be regarded as a linear expansion over all possible 16 linearly
independent basis operators operating on a four-dimensional Hilbert
space, with the expansion coefficients containing arbitrary bath
operators. Note however, the frequency spectrum of the bath is
assumed to have a hard cutoff so that the general theory of UDD
is applicable.

The basis operators used in the above-mentioned expansion can be taken as $
\{R_{i}\}_{i=1,2,\cdots,16}=\{\sigma_{k}\otimes\sigma_{l}\}$,
with
$\sigma_{k},\sigma_{l}\in\{I,\sigma_{x},\sigma_{y},\sigma_{z}\}$
($I$ the unity operator for the two-qubit Hilbert space) and the
orthogonality condition $\mathrm{Tr}(R_{j}R_{k})=4\delta_{jk}$.
This choice of basis operators is rather arbitrary. Purely for
the sake of discussions below, we find it convenient to define two
new sets of basis operators.
Let $|0\rangle$, $|1\rangle$, $|2\rangle$, and $|3\rangle$ be the four orthogonal basis states of the two-qubit Hilbert space, we define the following two new sets of basis operators,
\begin{equation}
\begin{aligned} Y_{1}=\tilde{Y}_{1} & =I,\\
Y_{2}=\tilde{Y}_{2} & =|0\rangle\langle0|+|1\rangle\langle1|,\\
Y_{3}=\tilde{Y}_{3} & =|2\rangle\langle2|-|3\rangle\langle3|,\\
Y_{4}=\tilde{Y}_{4} & =|2\rangle\langle3|,\\
Y_{5}=\tilde{Y}_{5} & =|3\rangle\langle2|,\\
Y_{6}=\tilde{Y}_{6} & =|0\rangle\langle0|-|1\rangle\langle1|, \\
Y_{7}=|1\rangle\langle2|; & \ \tilde{Y}_{7}=[|0\rangle\langle2|-|1\rangle\langle2|],\\
Y_{8}=|2\rangle\langle1|; & \ \ \tilde{Y}_{8}=[|2\rangle\langle0|-|2\rangle\langle1|],\\
Y_{9}=|1\rangle\langle3|; & \ \ \tilde{Y}_{9}=[|0\rangle\langle3|-|1\rangle\langle3|],\\
Y_{10}=|3\rangle\langle1|; & \ \ \tilde{Y}_{10}=[|3\rangle\langle0|-|3\rangle\langle1|],\\
Y_{11}=|0\rangle\langle2|; & \ \ \tilde{Y}_{11}=[|0\rangle\langle2|+|1\rangle\langle2|],\\
Y_{12}=|2\rangle\langle0|; & \ \ \tilde{Y}_{12}=[|2\rangle\langle0|+|2\rangle\langle1|],\\
Y_{13}=|0\rangle\langle3|; & \ \ \tilde{Y}_{13}=[|0\rangle\langle3|+|1\rangle\langle3|],\\
Y_{14}=|3\rangle\langle0|; & \ \ \tilde{Y}_{14}=[|3\rangle\langle0|+|3\rangle\langle1|],\\
Y_{15}=\tilde{Y}_{15} & =|0\rangle\langle1|+|1\rangle\langle 0|, \\
Y_{16}=\tilde{Y}_{16} & =-i(|1\rangle\langle0|-|0\rangle\langle
1|).\end{aligned} \label{ytilde}
\end{equation}
Then our general total Hamiltonian $H$ may be re-expressed as
\begin{eqnarray}
H=\sum_{i=1}^{16}W_{i}Y_{i}=\sum_{k=1}^{16}\tilde{W}_{k}\tilde{Y}_{k}, \label{Hform}
\end{eqnarray}
where $W_{i}$ or $\tilde{W}_{k}$ are the associated new expansion
coefficients containing bath operators. The motivation of using $Y_i$ ($i=1-16$) as the basis operators
is purely for convenience. For example, most of them have only one nonzero matrix element and some others are chosen to form
a standard SU(2) subalgebra.  Similarly, the basis operators $\tilde{Y}_i$ ($i=7-14$) are chosen to simplify our later calculations
involving states $(|0\rangle +|1\rangle)/\sqrt{2}$ [see Eq.~(\ref{phasec})].  Note also that
the new basis operators defined in Eq.~(\ref{ytilde}), which still form the generating algebra of $H$, may not take a Hermitian form.
This is not an issue because their linear superpositions still
generate all possible Hermitian operators for a two-qubit system.


\subsection{Locking a known two-qubit state by an extended UDD scheme}
Given a known but arbitrary two-qubit state, here assumed to be
$|0\rangle$ without loss of generality,
  we first construct a
control operator
\begin{eqnarray}
X_0=2|0\rangle\langle 0|-I,
\end{eqnarray} with $X_0^2=I$.  As recently pointed out in Ref.~\cite{Liureview}, such a control operator
was also considered in Ref. \cite{groverwork} before UDD was discovered.  We can now split $H$ into two parts,
\begin{eqnarray}
H&=&H_0+H_1, \nonumber\\
H_{0}&=&\sum_{i=1}^{10}W_{i}Y_{i}, \nonumber \\
H_1&=&\sum_{i=11}^{16}W_{i}Y_{i},
\end{eqnarray} with the commuting relation $[X_0,H_0]=0$, and the
anti-commuting relation $\{X_0,H_1\}_{+}=0$.  We proposed in
Ref.~\cite{pra1} the following control Hamiltonian describing a
sequence of extended UDD $\pi$-pulses over a duration of $T$, i.e.,
\begin{eqnarray}
\label{control1}
H_{c}=\sum_{j=1}^{N}\pi \delta(t-T_j)\frac{X_0}{2},
\label{UDDcontrol}
\end{eqnarray}
with the UDD timing $T_{j}$ given by
\begin{eqnarray} T_{j}  =T \sin^{2}(\frac{j\pi}{2N+2}),\
j=1, 2\cdots ,N.
\label{UDDtime}
\end{eqnarray}
For odd $N$, an additional control pulse is applied in the end.  Then the
unitary evolution operator for the whole system of two qubits in a bath for the
period $t=0$ to $t=T$ is given by ($\hbar=1$ throughout)
\begin{eqnarray}
{U}_{N}(T) & = & X_{0}^{N} e^{-i[H_{0}+H_1](T-T_N)} (-iX_0) \nonumber \\
& &\ \times\ e^{-i[H_{0}+H_1](T_N-T_{N-1})} (-iX_0) \nonumber \\
& &\ \cdots \nonumber \\
 & &\ \times\  e^{-i[H_{0}+H_1](T_3-T_2)} (-iX_0) \nonumber \\
& &\ \times\ e^{-i[H_{0}+H_1](T_2-T_1)} (-iX_0)\nonumber \\
& &\ \times\ \ e^{-i[H_{0}+H_1]T_1}. \label{Ut}
  \end{eqnarray}
 Exploiting  $[H_0,X_0]=0$ and $\{H_1,X_0\}_{+}=0$, one can
 directly use the UDD universality proof developed by Yang and Liu \cite{UDD3,Liureview}, yielding
 \begin{eqnarray}
 U_{N}(T)=U_{N}^{\text{even}}+O(T^{N+1}),
\end{eqnarray}
where
\begin{eqnarray}
U_N^{\text{even}}=\exp(-iH_0 T) \sum_{k=0}^{+\infty} (-i)^{2k}\Delta
_{2k}, \label{Deltaform}
\end{eqnarray}
with $\Delta_{2k}$ only containing even powers of $H_1^I(t)$, defined
by $H_1^I(t)\equiv \exp(iH_0 t) H_1\exp(-iH_0 t)$.

Because $\{H_1,X_0\}_{+}=0$, one has  $\{H_1^I(t),X_0\}_{+}=0$.   As
such, any even power of $H_1^I(t)$ will commute with $X_0$, e.g,
$[H_1^I(t_1)H_1^I(t_2),X_0]=H_{1}^{I}(t_1)\{H_{1}^{I}(t_2),X_0\}_{+}
- \{H_{1}^{I}(t_1),X_0\}_{+}H_{1}^{I}(t_2)=0$.  This important observation indicates
that $\Delta_{2k}$ can be expanded as a linear superposition of all
possible basis operators that commute with $X_0$. That is,
\begin{eqnarray}
\Delta_{2k}=\sum_{i=1}^{10} A_{i} Y_i,
\end{eqnarray}
where $A_i$ are the expansion coefficients containing bath operators.  Clearly then, to the $N$th order, $U_{N}(T)$
can be expressed as a combination of $Y_1$, $Y_2$, $\cdots$, $Y_{10}$ only. Using the closure of this set of operators, i.e.,
\begin{eqnarray}
\left(\sum_{i=1}^{10}A_{i}Y_{i}\right) \left(
\sum_{k=1}^{10}B_{k}Y_{k}\right)&=&\sum_{l=1}^{10}C_{l}Y_{l},
\end{eqnarray}
we further obtain
\begin{eqnarray}
U_{N}(T)= \exp(-iH_{\text{eff}}^{\text{UDD-1}} T)+ O(T^{N+1}),
\end{eqnarray}
where
\begin{eqnarray}
H_{\text{eff}}^{\text{UDD-1}} =  \sum_{i=1}^{10} D_{1,i} Y_i,
\end{eqnarray}
with $D_{1,i}$ being the expansion coefficients.

The outcome of applying a UDD sequence of $X_0$ is now evident by
comparing the original total Hamiltonian $H$ in the absence of
control with the effective Hamiltonian
$H_{\text{eff}}^{\text{UDD-1}}$ realized by UDD.  In essence, the
UDD sequence based on $X_0$ efficiently removes the operators
$Y_{11}$, $Y_{12}$, $\cdots$, $Y_{16}$ from the initial generating algebra of $H$, thus
suppressing all possible coupling between the pre-chosen
state $|0\rangle$ and all other states.  Therefore, given a known
entangled state $|0\rangle$, we can protect this state to the $N$th
order with $N$ [or $(N+1$)] instantaneous control pulses, a result
analogous to single-qubit UDD.

For later use, we list below three key requirements in achieving a
UDD-reduced effective Hamiltonian $H_{\text{eff}}^{\text{UDD-1}}$ to the $N$th order,
from a general Hamiltonian describing two qubits plus a bath:
\begin{enumerate}
\item[(i)] Construction of a control operator (e.g., $X_0$) whose square equals the unity operator.  This control operator will be used to form a UDD sequence of $N$ instantaneous pulses [e.g., Eq.~(\ref{UDDcontrol})].
\item[(ii)] Separation of the bare system-bath Hamiltonian into two terms, say $H_0$ and $H_1$, with
 $H_0$ commuting with the control operator and $H_1$ anti-commuting with the control operator.
\item[(iii)] Algebra closure of the operators forming $H_0$, which becomes the generating algebra of a UDD-reduced effective Hamiltonian.
\end{enumerate}
In our following considerations we will make a number of references to these three requirements.

\section{Protecting unknown two-qubit entangled states by nested UDD}
\subsection{First scheme for nesting three UDD layers}
The $X_0$ control operator in Sec. II is based on the knowledge that
the state to be preserved is $|0\rangle$.  It is even more useful if
we can develop a scheme to protect unknown two-qubit entangled
states. To that end, let us assume that an unknown two-qubit state
$|\psi(0)\rangle$ is a superposition of two orthogonal basis states,
e.g., $|0\rangle$ and $|1\rangle$.  Though our considerations below
are general, to be specific we assume $|0\rangle\equiv
|\uparrow\uparrow\rangle$, and $|1\rangle\equiv
|\downarrow\downarrow\rangle$, where $\uparrow$ and $\downarrow$
represent spin-up and spin-down states of each qubit. The unknown
two-qubit state to be protected can be written as
\begin{eqnarray}
 |\psi(0)\rangle&=&\alpha|0\rangle + \beta |1\rangle \nonumber \\
&=& \alpha|\uparrow\uparrow\rangle + \beta |\downarrow\downarrow\rangle,
\end{eqnarray}
where the two unknown coefficients $\alpha$ and $\beta$ satisfy
$|\alpha|^2+|\beta|^2=1$ at time zero. Can we efficiently protect
such type of unknown entangled states by further extending UDD?  Note that this problem is different
from a single-qubit case because the population can leak out from the initial two-dimensional subspace.

We use $\rho(t)$ to represent the total density matrix of the system
and the bath at time $t$, evolving from a direct product state of
$|\psi(0)\rangle$ and some initial state of the bath. The protection
of the state $|\psi(0)\rangle$ requires to freeze multiple coherence
properties, i.e., diagonal populations
$\text{Tr}[\rho(t)|\uparrow\uparrow\rangle\langle
\uparrow\uparrow|]\approx |\alpha|^2 $ and
$\text{Tr}[\rho(t)|\downarrow\downarrow\rangle\langle
\downarrow\downarrow|]\approx |\beta|^2$, as well as the
off-diagonal phase property
$\text{Tr}[\rho(t)|\uparrow\uparrow\rangle\langle
\downarrow\downarrow|] \approx \alpha \beta^{*}$.  This  motivates
us to extend the UDD nesting scheme in Ref.~\cite{west}, where both
single-qubit population relaxation and single-qubit transverse
dephasing are suppressed in a near-optimal fashion. Certainly, our
problem here is more demanding: in single-qubit systems with a
two-dimensional Hilbert space, the locking of one projection
probability onto one basis state automatically freezes the
projection probability onto a second basis state, whereas here the
locking of the diagonal probabilities at $|\alpha|^2$ and $|\beta|^2$
should be respectively achieved by control pulses. Given that two
layers of UDD sequences are needed for complete single-qubit
decoherence control in Ref.~\cite{west}, it is a natural guess that
we will at least need three layers of nested UDD sequences.

We can now directly make use of our results in the previous section.
In the first step, we lock the diagonal property $\text{Tr}
[\rho(t)|\uparrow\uparrow\rangle\langle \uparrow\uparrow|]$. This
can be achieved by considering an innermost layer of UDD sequence of
$X_0$, such that all possible coupling between $|0\rangle$ and all
other orthogonal states can be efficiently removed. Doing so, we
reduce $H$ to $H_{\text{eff}}^{\text{UDD-1}}$ to the $N$th order, as
elaborated in Sec. II.  The population on state $|0\rangle$ is hence
locked.

In the second step, we treat a decoherence control problem for the
effective Hamiltonian $H_{\text{eff}}^{\text{UDD-1}}$.  We assume
that $H_{\text{eff}}^{\text{UDD-1}}$ resulting from the innermost
layer of UDD is a sufficiently smooth function of time, such that a
second layer of UDD can be applied to
$H_{\text{eff}}^{\text{UDD-1}}$.  Note however, though this is a
natural assumption and Ref. \cite{uhrigJA} reasoned about
the smoothness of analogous UDD-reduced effective Hamiltonians, some
counter examples might exist \cite{Liureview}.  With this smoothness assumption
exercised with caution,  we now introduce a second UDD layer to
lock the second diagonal property $\text{Tr}
[\rho(t)|\downarrow\downarrow\rangle\langle \downarrow\downarrow|]$.
As is clear from Sec. II,  in order to remove all possible couplings between
$|1\rangle$ and all other states, one may apply
the control operator $X_{1}=2|1\rangle\langle1|-I$ with
$X_1^2=I$.  To examine if this is
feasible, we decompose $H_{\text{eff}}^{\text{UDD-1}}$ into two
terms, i.e.,
\begin{eqnarray}
H_{\text{eff}}^{\text{UDD-1}} &= & \sum_{i=1}^{10} D_{1,i} Y_i \nonumber \\
&= & H_{\text{eff,0}}^{\text{UDD-1}}+
H_{\text{eff,1}}^{\text{UDD-1}},
\end{eqnarray}
with \begin{eqnarray}
 H_{\text{eff,0}}^{\text{UDD-1}} &\equiv & \sum_{i=1}^{6}D_{1,i} Y_i;  \nonumber \\
 H_{\text{eff,1}}^{\text{UDD-1}} &\equiv &  \sum_{i=7}^{10}D_{1,i} Y_i.
 \end{eqnarray}
Because $[H_{\text{eff,0}}^{\text{UDD-1}},X_1]=0$ and
$\{H_{\text{eff,1}}^{\text{UDD-1}},X_1\}_+=0$, it is seen that
$H_{\text{eff}}^{\text{UDD-1}}$ and $X_1$ guarantee requirement (ii)
outlined in Sec. II.
 Finally, it is straightforward to see that
the operators $Y_i$, $i=1-6$ form a closed algebra
[requirement (iii) above].  We hence expect that if a second layer
of UDD sequence of $X_1$ is applied, to the $N$th order
the dynamics of $H_{\text{eff}}^{\text{UDD-1}}$ under the second control layer  becomes that of a
simpler effective Hamiltonian (denoted $H_{\text{eff}}^{\text{UDD-2}}$) generated by a further
reduced algebra.  That is,
\begin{eqnarray}
H_{\text{eff}}^{\text{UDD-2}}=\sum_{i=1}^{6}D_{2,i} Y_i,
\end{eqnarray}
where $D_{2,i}$ represent the expansion coefficients due to two UDD layers.

Examining the self-closed set of operators that form  $H_{\text{eff}}^{\text{UDD-2}}$, one sees that only the component $D_{2,6}Y_6=D_{2,6}[|0\rangle\langle0|-|1\rangle\langle 1|]$ can affect the intial superposition state $\alpha|0\rangle + \beta |1\rangle$. Further,
this $Y_6$ component does not change the populations on states $|0\rangle$ and $|1\rangle$, so it represents a pure dephasing mechanism. To efficiently suppress this pure dephasing, we now consider a third, outermost UDD layer.  Assuming again that $H_{\text{eff}}^{\text{UDD-2}}$ is sufficiently smooth for UDD to apply, we introduce the following ``phase" control operator
\begin{eqnarray}
X_{\phi}\equiv [|0\rangle+|1\rangle][\langle0|+\langle 1|]-I,
\label{phasec}
\end{eqnarray}
with $X_\phi^2=I$.  Separating $H_{\text{eff}}^{\text{UDD-2}}$ into two terms, we obtain
\begin{eqnarray}
H_{\text{eff}}^{\text{UDD-2}} = H_{\text{eff},0}^{\text{UDD-2}}+H_{\text{eff},1}^{\text{UDD-2}},
\end{eqnarray}
with
\begin{eqnarray}
H_{\text{eff},0}^{\text{UDD-2}}& = & \sum_{i=1}^{5}D_{2,i} Y_i;\nonumber \\
H_{\text{eff},1}^{\text{UDD-2}}& = & D_{2,6} Y_6.
\end{eqnarray}
Interestingly, $[H_{\text{eff},0}^{\text{UDD-2}},X_\phi]=0$ and
$\{H_{\text{eff},1}^{\text{UDD-2}},X_{\phi}\}_{+}=0$.  Further, the
operators $Y_i$, $i=1-5$ form a closed
algebra.  All the three key requirements for UDD are again met for this outermost layer. The
final reduced effective Hamiltonian after three layers of UDD is
hence formed by five operators, i.e.,
\begin{eqnarray}
H_{\text{eff}}^{\text{UDD-3}}=\sum_{i=1}^{5} D_{3,i}Y_i,
\end{eqnarray}
where $D_{3,i}$ represent the expansion coefficients due to three UDD layers.  Referring to the subspace spanned by $|0\rangle$ and
$|1\rangle$, $H_{\text{eff}}^{\text{UDD-3}}$ only contains an
identity operator for that subspace. Hence any unknown initial
superposition state $\alpha|0\rangle + \beta |1\rangle$ is well
preserved to the $N$th order.

In terms of step-by-step reduction of the generating algebra associated with the effective Hamiltonians at each level,
the following chart summarizes how our nesting scheme reduces a general total Hamiltonian to a much simplified
form:
\vspace{0.3cm}
\begin{center}
\framebox{$Y_i$, $i=1,2, \cdots, 16$} \\
\vspace{0.2cm}
$\Downarrow${$X_0$, \text{UDD-1}}\\
\vspace{0.2cm}
\framebox{$Y_i$, $i=1,2, \cdots, 10$} \\
\vspace{0.2cm}
$\Downarrow${$X_1$,\text{UDD-2}}\\
\vspace{0.2cm}
\framebox{$Y_i$, $i=1,2,\cdots, 6$}\\
\vspace{0.2cm}
$\Downarrow${$X_\phi$,\text{UDD-3}}\\
\vspace{0.2cm}
\framebox{$Y_i$, $i=1,2, \cdots, 5$}.
\end{center}
The operators inside each box represent the elements of the
generating algebra before or after a certain layer of UDD.

For completeness, we also explicitly present here the timing of the
control pulses within each UDD layer. In particular, the control
operator $X_0$ in the innermost layer is applied at
\begin{equation}
T_{j,k,l}  =
T_{j,k}+(T_{j,k+1}-T_{j,k})\sin^{2}\left(\frac{l\pi}{2N+2}\right),
\end{equation}
where $T_{j,k}$ is the UDD timing for $X_1$ in the middle layer.  $T_{j,k}$ is given by
\begin{equation}
T_{j,k}
=T_{j}+(T_{j+1}-T_{j})\sin^{2}\left(\frac{k\pi}{2N+2}\right),
\end{equation}
where $T_j$ represents the timing for $X_\phi$ in the outermost layer and it is
already given by Eq.~(\ref{UDDtime}).  Similar to single-qubit
cases, for each layer, if $N$ is odd then an additional control
operator is applied at the end of each sequence. Overall, $N^3$ [or
$(N+1)^3$] control pulses are applied to achieve decoherence
suppression to the $N$th order.  Because states $|0\rangle$ and
$|1\rangle$ play a similar role here, our analysis above equally
applies if $X_1$ and $X_0$ are exchanged.

 To realize such a UDD nesting scheme with this high-order decoherence suppression,
 the involved two-qubit control operators are nonlocal control operators in general.  Taking $|0\rangle=|\uparrow\uparrow\rangle$ and $|1\rangle=|\downarrow\downarrow\rangle$ as an example, Table I lists the three control operators (another operator to be explained later) in terms of the familiar Pauli matrices. The identity operator in the expressions for $X_0$, $X_1$ and $X_\phi$ in Table I is not important.   Experimentally, realizing such two-qubit control operators is analogous to realizing quantum computation in a two-qubit system. The true challenge might lie in realizing a sufficient speed of such two-qubit operations.

\begin{table}
\caption{Explicit construction of control operators for nesting three layers of UDD sequences in order to protect unknown two-qubit entangled states $\alpha|0\rangle+\beta |1\rangle$, with
$|0\rangle=|\uparrow\uparrow\rangle$ and $|1\rangle=|\downarrow\downarrow\rangle$. \\ }
\centering
\begin{tabular}{|c|c|}
\hline
Constructed\ Operator & Explicit\ Form \\ \hline
$X_0$ & $\left(\sigma_z^{1}\sigma_z^2+\sigma_z^1+\sigma_z^2-I\right)/2$ \\
\hline $X_1$& $
\left(\sigma_z^{1}\sigma_z^2-\sigma_z^1-\sigma_z^2-I\right)/2$ \\
\hline $X_\phi$ &
$\left(\sigma_x^{1}\sigma_x^2-\sigma_y^1\sigma_y^2+\sigma_z^1\sigma_z^2-I\right)/2$\\
\hline $X_{0,1}$ & $\sigma_z^1\sigma_z^2$\\ \hline
\end{tabular}
\end{table}

\subsection{Wrong ordering of three UDD layers}
At this point an interesting question arises. That is, does the
ordering of the three nested UDD sequences matter or not? To answer
this question let us first exchange the ordering of the two
sequences of $X_0$ and $X_\phi$, such that $X_\phi$ is in the
innermost layer and $X_0$ is in the outermost layer.  We denote this ordering
as $X_0-X_{1}-X_{\phi}$. In the following we shall stick to this convention for ordering, i.e., an operator appearing at the
rightmost (leftmost) will be placed in the innermost (outermost) layer.
Because the
$X_\phi$ layer is now operating directly on the bare Hamiltonian
$H$, we re-partition $H$ in Eq.~(\ref{Hform}) into the following two
terms, i.e.,
\begin{eqnarray}
H&=& \left(\sum_{k=1}^{5}\tilde{W}_{k} \tilde{Y}_{k}+\sum_{k=7}^{10} \tilde{W}_{k} \tilde{Y}_{k}+ \tilde{W}_{15} \tilde{Y}_{15}\right) \nonumber \\
 && + \left(\tilde{W}_{6} \tilde{Y}_{6}+\tilde{W}_{16} \tilde{Y}_{16}+\sum_{k=11}^{14}\tilde{W}_{k} \tilde{Y}_{k}\right),
\end{eqnarray}
where operators $\tilde{Y}_k$ are defined in Eq.~(\ref{ytilde}). The
first term in the above equation commutes with $X_\phi$, whereas the
second term anti-commutes with $X_\phi$. All the operators contained
in the first term form a closed algebra. The innermost UDD sequence
of $X_\phi$ hences yields an effective Hamiltonian
\begin{eqnarray}
\tilde{H}_{\text{eff}}^{\text{UDD-1}} &= & \sum_{i=1}^{5} \tilde{D}_{1,i}\tilde{Y}_i  + \sum_{i=7}^{10} \tilde{D}_{1,i}  \tilde{Y}_i + \tilde{D}_{1,15}  \tilde{Y}_{15}\nonumber \\
 \end{eqnarray}
where $\tilde{D}_{1,i}$ are the expansion coefficients.  To consider the
second UDD layer, we rewrite $\tilde{H}_{\text{eff}}^{\text{UDD-1}}$
in terms of $Y_{k}$, i.e.,
\begin{eqnarray}
\tilde{H}_{\text{eff}}^{\text{UDD-1}} &= & \left(\sum_{i=1}^{5} D_{1,i}' {Y}_i  + \sum_{i=11}^{14} D_{1,i}'{Y}_i\right) \nonumber \\
&& + \left(\sum_{i=7}^{10} D_{1,i}' {Y}_i  + D_{1,15}' {Y}_{15}\right),
\label{newHeff1}
\end{eqnarray}
where $D_{1,i}'$ is connected with $\tilde{D}_{1,k}$ via a simple relation
between $Y_i$ and $\tilde{Y}_i$ [see Eq.~(\ref{ytilde})].

The middle layer of UDD sequence would be based on the control
operator $X_1$. Among those basis operators that form
$\tilde{H}_{\text{eff}}^{\text{UDD-1}}$, $X_1$ commutes with those in the
first line of Eq.~(\ref{newHeff1}) and anti-commutes with those in the
second line of Eq.~(\ref{newHeff1}).  This indicates that the final
evolution operator associated with this UDD layer can be cast into a
form similar to Eq.~(\ref{Deltaform}).  We next investigate if the
key requirement (iii) of UDD can be satisfied. Interestingly and
somewhat unexpectedly, this is not the case: operators $Y_i$ with
$i=1,2, \cdots, 5$ together with $Y_{i}$ with $i=11-14$ cannot form
a closed algebra.  For example, $2Y_{11}Y_{12}-Y_2$ will yield
$Y_6$, which is already outside this collection of basis operators.
As a result, the application of this second control layer will not yield
a further reduced effective Hamiltonian. Such a nesting scheme then
breaks down due to its wrong ordering! Indeed, in the correctly
ordered case, the dephasing operator $Y_6$ will be suppressed by
$X_\phi$, but here it resurfaces (from even powers of those
operators that commute with $X_1$) after the $X_\phi$ layer is
already applied.

Note however, due to this wrong ordering, the undesired operators (such
as $Y_6$ that cannot be suppressed by the outermost $X_0$ layer)
reemerge from multiplications of a set of basis operators. We hence
intuitively expect its impact on decoherence control to be at least a second-order
effect. We will come back to this when discussing our
numerical results.

One may wonder what happens to other ordering?  Because states
$|0\rangle$ and $|1\rangle$ play the same role here, there is only
one non-equivalent ordering left, i.e., $X_{1}-X_\phi-X_{0}$ (or
equivalently, $X_{0}-X_\phi-X_{1}$).
 The effective Hamiltonian
after the innermost layer of UDD sequence is hence still given by
$H_{\text{eff}}^{\text{UDD-1}}=\sum_{i=1}^{10} D_{1,i} Y_i$. Since
the next layer of control is $X_\phi$, one checks if
$H_{\text{eff}}^{\text{UDD-1}}$ can be decomposed into two terms in
accord with the properties of $X_\phi$ [to fulfill requirement (ii)
outlined above]. Interestingly, though this procedure can be easily done in the entire two-qubit operator space,
it cannot be done here within the reduced generating algebra of  $H_{\text{eff}}^{\text{UDD-1}}$.
That is, among the set of basis operators that form
$H_{\text{eff}}^{\text{UDD-1}}$, some operators or their arbitrary combinations are neither commuting nor
anti-commuting with $X_\phi$. For example,
\begin{eqnarray}
[Y_7,X_\phi]\ne 0;\ \ \{Y_7,X_\phi\}_{+}\ne 0.
\end{eqnarray}
So even without checking if there is a self-closed set of operators
that commute with $X_\phi$, it is already seen that the nesting scheme breaks
down directly in this ordering.    Thus, out of three non-equivalent
ordering of the three control operators considered here, only the
ordering advocated in the previous subsection may achieve
high-order protection of two-qubit states.

From a more general perspective, the dependence of the controlled dynamics on
the ordering of the three UDD layers can be explained as follows. Consider a particular time interval
$\Delta T$ for a time-independent Hamiltonian $H$, during which each of the two control operators $A$ and $B$ ($A^2=1$ and $B^2=1$) are applied twice with a certain ordering. If $B$ is nested inside, the associated unitary evolution is given by
\begin{equation}
U_{A-B}=AB e^{-iH\Delta T}BA= e^{-i\left[(AB)H (BA)\right]\Delta T};
\label{eq1-a}
\end{equation} whereas for the other ordering, the unitary evolution is given by
\begin{equation}
U_{B-A}=
BA e^{-iH\Delta T}AB  =e^{-i\left[(BA)H(AB)\right]\Delta T}.
\label{eq2-a}
\end{equation}
As seen from the right hand side of Eqs.~(\ref{eq1-a}) and ~(\ref{eq2-a}), the two ordering leads to
two effective Hamiltonians $(AB)H (BA)$ and $(BA)H(AB)$.  In general these two effective Hamiltonians are different, thus giving rise to
the ordering dependence.  Note however, if for a concerned Hilbert subspace $AB=BA$ or $AB=-BA$, i.e., if the two control operators commute or anti-commute, then
these two effective Hamiltonians are identical and hence the ordering dependence no longer exists.
This further explains why we cannot exchange the ordering between $X_\phi$ and $X_0$ or the ordering between $X_\phi$ and $X_1$, but can exchange the ordering between $X_0$ and $X_1$.

\subsection{Alternative Nesting Scheme}
For the two-dimensional subspace spanned by states $|0\rangle$ and $|1\rangle$,  any two orthogonal
 states $|0'\rangle$ and $|1'\rangle$ can be adopted to construct analogous
  control operators $X_{0'}$, $X_{1'}$, and $X_{\phi'}$ for three UDD layers. Therefore,
  when it comes to an actual implementation, there are infinite possibilities to realize three nested UDD layers in order
    to protect an unknown entangled state in that subspace.

Even for fixed basis states $|0\rangle$ and $|1\rangle$, the nesting scheme with the right ordering
analyzed above is not the only solution.  Consider the following control operator
\begin{eqnarray}
{X}_{0,1}= 2[|0\rangle\langle 0|+|1\rangle \langle 1|]-I= X_{0}+X_1+I
\end{eqnarray}
with $X_{0,1}^2=I$. The explicit form of $X_{0,1}$ is also given in
Table I. Following the analysis in Sec. II,  it is straightforward
to confirm that a UDD sequence of $X_{0,1}$ can efficiently freeze
the total population in the two-dimensional subspace spanned by
$|0\rangle$ and $|1\rangle$.   We can now construct an alternative nesting scheme using $X_{0,1}$, $X_\phi$ and
$X_1$ (or $X_0$).

The ordering of $X_\phi-X_1-X_{0,1}$ is studied first.  It is found that
each of the three UDD layers satisfies the three requirement outlined in Sec. II and hence yields a simple effective
Hamiltonian to the $N$th order. Qualitatively, it is also obvious
why this scheme is expected to work. The innermost layer effectively
lock the population in a two-dimenional Hilbert subspace. Then, in
essence, the next two layers are similar to those in single-qubit
two-layer UDD~\cite{west}, insofar as a second layer locks
one population and the outermost layer freezes the relative phase
between the two projection amplitudes.  Extending this analogy, it
is expected that the ordering of the middle layer and the outermost
layer can be exchanged (indeed, the associated control operators anti-commute in the two-dimensional subspace).
This can be more formally analyzed by
working out the detailed algebra layer-by-layer. In particular, for
the ordering of $X_\phi-X_1-X_{0,1}$, the generating algebra
for the effective Hamiltonians after each control layer reduces in
the following fashion: \vspace{0.3cm}
\begin{center}
\framebox{$Y_i$, $i=1,2, \cdots, 16$} \\
\vspace{0.2cm}
$\Downarrow${$X_{0,1}$, \text{UDD-1}}\\
\vspace{0.2cm}
\framebox{$Y_i$, $i=1,2, \cdots, 6$; $Y_{15}$, $Y_{16}$} \\
\vspace{0.2cm}
$\Downarrow${$X_1$,\text{UDD-2}}\\
\vspace{0.2cm}
\framebox{$Y_i$, $i=1,2,\cdots, 6$}\\
\vspace{0.2cm}
$\Downarrow${$X_\phi$,\text{UDD-3}}\\
\vspace{0.2cm}
\framebox{$Y_i$, $i=1,2, \cdots, 5$}.
\end{center}
\vspace{0.3cm}
For the ordering of  $X_1-X_\phi-X_{0,1}$, we have
\vspace{0.3cm}
\begin{center}
\framebox{$Y_i$, $i=1,2, \cdots, 16$} \\
\vspace{0.2cm}
$\Downarrow${$X_{0,1}$, \text{UDD-1}}\\
\vspace{0.2cm}
\framebox{$Y_i$, $i=1,2, \cdots, 6$; $Y_{15}$, $Y_{16}$} \\
\vspace{0.2cm}
$\Downarrow${$X_\phi$,\text{UDD-2}}\\
\vspace{0.2cm}
\framebox{$Y_i$, $i=1,2,\cdots, 5$; $Y_{15}$}\\
\vspace{0.2cm}
$\Downarrow${$X_1$,\text{UDD-3}}\\
\vspace{0.2cm}
\framebox{$Y_i$, $i=1,2, \cdots, 5$}.
\end{center}
\vspace{0.3cm}
As seen above, both nesting strategies are successful and in the end the same generating algebra for the final effective Hamiltonian is reached.

Can we place a sequence of $X_{0,1}$ in the middle layer instead?  For the ordering of
$X_\phi-X_{0,1}-X_1$, since the population on state $|1\rangle$ is already locked by the inner most layer,
the role of the second layer is equivalent to further locking the population on state $|0\rangle$,
thus playing a similar role as $X_{0}$. The outermost layer then freezes the relative phase between
the two projection amplitudes, analogous to the case of  $X_\phi-X_{0}-X_1$.  In this sense, the ordering of
 $X_\phi-X_{0,1}-X_1$ does not provide anything new.  Indeed, based on the discussion at the end of the last subsection,
 because $[X_{0,1},X_1]=0$, their ordering is expected to be exchangeable.

Consider then the other ordering $X_1-X_{0,1}-X_\phi$ (which is also equivalent to $X_0-X_{0,1}-X_\phi$).
In this case, the effective Hamiltonian
reduced by the innermost layer is given by $\tilde{H}_{\text{eff}}^{\text{UDD-1}}$ in  Eq.~(\ref{newHeff1}).  To analyze the effect of the second layer, we re-split $\tilde{H}_{\text{eff}}^{\text{UDD-1}}$ in Eq.~(\ref{newHeff1}) as
\begin{eqnarray}
\tilde{H}_{\text{eff}}^{\text{UDD-1}} &= & \left(\sum_{i=1}^{5} D_{1,i}' {Y}_i  +  D_{1,15}' {Y}_{15}\right) \nonumber \\
&&+ \sum_{i=7}^{14} D_{1,i}'{Y}_i,
\label{newHeff1-newform}
\end{eqnarray}
where the first term commutes with $X_{0,1}$ and the second term anti-commutes with $X_{0,1}$.  As an interesting outcome, now the set of operators that commute with the second-layer control operator also form a closed algebra, thus paving the way for the third UDD layer. Indeed, the outermost $X_0$ layer further reduces the algebra by removing the $Y_{15}$ component and hence yields an effective Hamiltonian seen before.  Summarizing, the explicit algebra reduction route for $X_1-X_{0,1}-X_\phi$ becomes
\vspace{0.3cm}
\begin{center}
\framebox{$Y_i$, $i=1,2, \cdots, 16$} \\
\vspace{0.2cm}
$\Downarrow${$X_{\phi}$, \text{UDD-1}}\\
\vspace{0.2cm}
\framebox{$Y_i$, $i=1, \cdots, 5$; $Y_i$, $i=7,8, \cdots, 15$} \\
\vspace{0.2cm}
$\Downarrow${$X_{0,1}$,\text{UDD-2}}\\
\vspace{0.2cm}
\framebox{$Y_i$, $i=1,2,\cdots, 5$; $Y_{15}$}\\
\vspace{0.2cm}
$\Downarrow${$X_1$,\text{UDD-3}}\\
\vspace{0.2cm}
\framebox{$Y_i$, $i=1,2, \cdots, 5$}.
\end{center}
\vspace{0.3cm}

We have also examined what happens if the $X_{0,1}$ layer is placed
in the outermost layer. For reasons
analogous to our first nesting scheme using $X_0$, $X_1$ and $X_\phi$, such type of ordering cannot
simplify the effective Hamiltonians layer-by-layer.  This concludes this subsection.

\subsection{Numerical study}
Similar to our previous work~\cite{pra1}, we use a five-spin system
to carry out simple numerical experiments.  Two of the five spins
are identified as our two-qubit system, and the other three spins
are regarded as the bath.   To avoid assumptions about how the
system is coupled to the bath, we work with the following general
total Hamiltonian
\begin{eqnarray}
\label{SBH}
H & = & \sum_{m=1}^{5}\sum_{\gamma=\{x,y,z\}}b_{\gamma,m}\sigma_{\gamma}^{m}\nonumber \\
 &  & + \sum_{m=1}^{5}\sum_{\gamma=\{x,y,z\}}\sum_{n>m}^{5}\sum_{\delta=\{x,y,z\}}c_{\gamma\delta}^{mn}\sigma_{\gamma}^{m}\sigma_{\delta}^{n}
\end{eqnarray} in dimensionless units, where all the coefficients $b_{\gamma,m}$ and
$c_{\gamma\delta}^{mn}$ are randomly sampled from the range $[-0.5,0.5]$.
We average our results over ten random realizations of this five-spin
system-bath Hamiltonian. In addition, to demonstrate that our approach
does not depend on the actual form of an initial superposition state
$\alpha|\uparrow\uparrow\rangle+ \beta|\downarrow\downarrow\rangle$, we
further average our results over ten initial states with randomly
sampled coefficients $\alpha$ and $\beta$ under the constraint
$|\alpha|^2+|\beta|^2=1$.

\begin{figure}[t]
\begin{center}
\hspace{-1cm}
\resizebox*{9.5cm}{8cm}{\includegraphics*{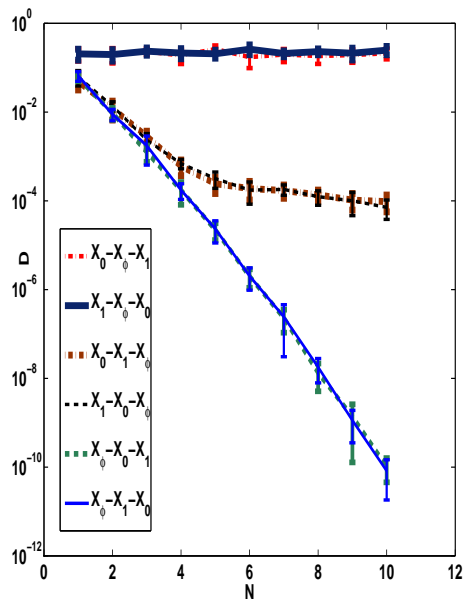}}
\end{center}
 \caption{(color online) Averaged trace distance $D$ between initial and final two-qubit states in the presence of three-layer nested UDD.
 $N$ is the number of
 instantaneous control pulses for each layer. The three control operators $X_{0}$, $X_{1}$, and $X_\phi$ are explicitly constructed in Table I.
 The total system-bath Hamiltonian is modeled by Eq.~(\ref{SBH}). Note a strong dependence on the ordering of the three UDD layers.  Top two curves are for $X_0-X_\phi-X_1$ and $X_1-X_\phi-X_0$; bottom two curves are for $X_\phi-X_0-X_1$ and $X_\phi-X_1-X_0$.
 All the variables plotted here and in all other figures are in dimensionless units.}
\end{figure}

Figure 1 depicts the averaged trace distance (denoted $D$) between the
system's reduced density matrix at time $t=0.1$ and its initial
state, for $N=1-10$. Different ordering of three UDD
sequences $X_0$, $X_1$, and $X_\phi$ are plotted together
for comparison. Consider first the two cases (bottom two curves) with the correct
ordering, i.e., $X_\phi-X_1-X_0$ and $X_\phi-X_0-X_1$.  For these
two cases a remarkably high fidelity is achieved in locking the
initial unknown superposition state. For $N=10$ (totally $N^3$ UDD
pulses), $D$ already reaches the $10^{-10}$ level.  The almost linear
scaling of $\log(D)$ vs $N$ is consistent with the expectation that
the extent of the decoherence suppression for a working nesting
scheme is to the $N$th order.

Turning to the top two flat curves associated with the wrong ordering
$X_1-X_\phi-X_0$ and $X_0-X_\phi-X_1$.  Their $D$ values do not
decrease with $N$ and stay at about $10^{-1}$. Therefore, for these
two cases the three-layer nested UDD does not work at all due to the
wrong ordering. This directly confirms our early insights into the
issue. In particular, from the algebra considerations we observe
that for the incorrect ordering here, the second layer of UDD
directly breaks down because the effective Hamiltonian reduced from
the inner most layer does not meet requirement (ii).

Still referring to Fig. 1, let us now discuss the middle curves
associated with another type of wrong ordering, i.e.,
$X_1-X_0-X_\phi$ and $X_0-X_1-X_\phi$.  It is seen that their $D$ values first decrease and then tend to saturate as $N$ increases.
The smallest $D$ values for $N=10$ is about $10^{-4}$, which is about six orders
of magnitude larger than in previous correctly ordered cases.
However, this performance is at the same time better than the
top two flat curves.  As such, the wrong ordering here
represents a weak deviation from an ideal nesting. This is
consistent with our early intuition that for the current ordering,
the UDD nesting scheme breaks down due to a high-order effect, i.e.,
the non-closure of a set of operators that commute with the control operator in the middle layer.

\vspace{0.3cm}
\begin{figure}[t]
\begin{center}
\vspace*{0.3cm}
\hspace{-1cm}
\par
\resizebox*{9.5cm}{8cm}{\includegraphics*{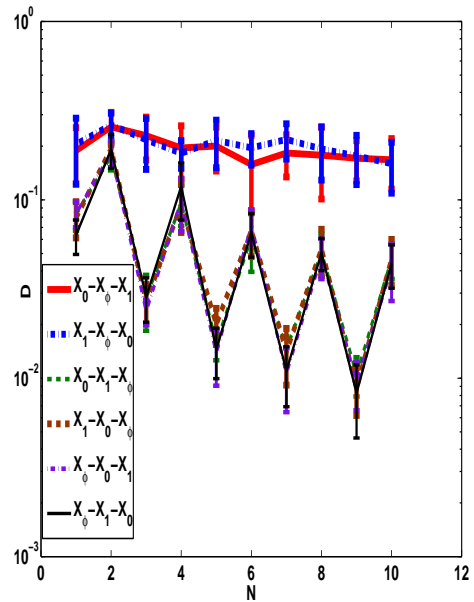}}
\end{center}
\par
 \caption{(color online) Same as in Fig. 1, except that the control pulses are applied periodically within each control layer. Note that the scale of the plotted $D$ values is many orders of magnitude different from that in Fig. 1.}
\end{figure}

As a comparison with a conventional dynamical decoupling approach
based on control pulses equally spaced in time, Fig. 2 presents the
parallel results if, within each layer, the control operator is
applied periodically.  Clearly, in this case, irrespective of the
ordering of the control operators, the performance of decoherence
control for $N=10$ in all cases is about nine orders of magnitude
worse than the best two cases in Fig. 1.  It is also observed that
the $D$ values are very weakly dependent on $N$. Results here remind
us that in addition to the ordering of the three layers, the timing
of the control operators is essential.

\hspace{-1cm}
\begin{figure}[t]
\begin{center}
\hspace{-1cm}
\par
\resizebox*{9.5cm}{8cm}{\includegraphics*{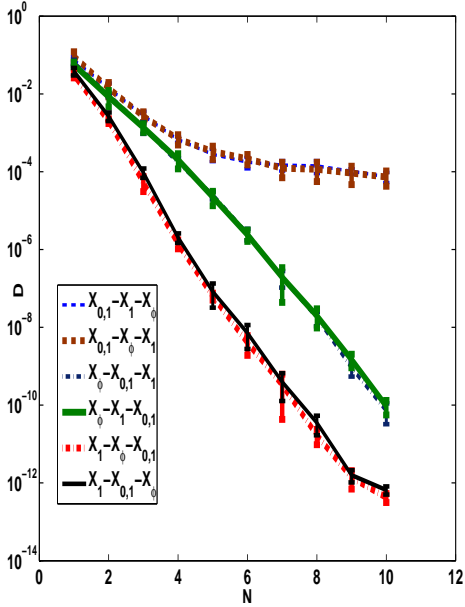}}
\end{center}
\par
 \caption{(color online) Same as in Fig. 1, except that the three control operators used in the three UDD layers are now $X_{0,1}$, $X_{1}$, and $X_\phi$. Top two curves are for $X_{0,1}-X_1-X_\phi$ and $X_{0,1}-X_\phi-X_{1}$; bottom two curves are for  $X_{1}-X_\phi-X_{0,1}$ and $X_{1}-X_{0,1}-X_\phi$.
 }
\end{figure}

Finally, results for an alternative UDD nesting scheme based on
$X_{0,1}$, $X_{1}$, and $X_{\phi}$ are shown in Fig. 3. The top two
curves are for two incorrect ordering $X_{0,1}-X_{1}-X_{\phi}$ and
$X_{0,1}-X_{\phi}-X_{1}$, with their $D$ values
saturating at about $10^{-4}$ as $N$ increases. It is noted that
the $X_{0,1}-X_{1}-X_{\phi}$ curve is similar to the $X_{0}-X_{1}-X_{\phi}$ case in Fig. 1.  This is understandable because the underlying
mechanism for unsuccessful nesting is the same. However, the $X_{0,1}-X_{\phi}-X_{1}$ case here has better performance than the $X_{0}-X_{\phi}-X_{1}$ case in Fig. 1. This is interesting because in both cases, the effective Hamiltonian reduced from the innermost layer does not satisfy requirement (ii) for the second layer.  Clearly then, for the outermost layer, a sequence of $X_{0,1}$ turns out to be superior to a sequence of $X_{0}$. This is somewhat expected
because $X_{0,1}$ in the outermost layer can still freeze the total population in the two-dimensional subspace whereas $X_{0}$ cannot.

All other four curves in Fig. 3 display a roughly linear scaling of
$\log(D)$ vs $N$, indicating the success of three-layer nested UDD.
Indeed, the layer ordering associated with these curves is
all predicted to be correct in our theoretical analysis above.
Interestingly, for fixed $N$,  the performance for the ordering of
$X_{\phi}-X_{1}-X_{0,1}$ or $X_{\phi}-X_{0,1}-X_{1}$ can differ from
that for $X_{1}-X_{\phi}-X_{0,1}$ or $X_{1}-X_{0,1}-X_{\phi}$ by
about two orders of magnitude. Comparing with the best
performance here with that in Fig. 1, a difference about two orders
of magnitude is also observed. These numerical details indicate that even with
the same scaling with $N$, the actual performance of a correctly ordered three-layer UDD
may depend on the specific algebra
reduction route.  More insights into this intriguing finding might help to further
understand the nature of decoherence dynamics under nested
multi-layer UDD.

\section{From three-layer UDD to four-layer UDD}

We are optimistic that for $N\sim 10$ a total of $N^3$ UDD control pulses as proposed in this work might be achievable in some systems in the near future.
If this is the case, then as the next step one wonders if there exist even better schemes, at least in theory.
In particular, as a result of three correctly ordered UDD layers,  the generating algebra for the final effective Hamiltonian $H_{\text{eff}}^{\text{UDD-3}}$ contains only five operators.  Can we construct better control operators to reduce the algebra more rapidly? Can we even consider one more UDD layer to remove all possible system-environment coupling?

 As a brief summary of our latest progress along these two questions, we first note that, the two nesting schemes proposed in Sec. III treat the subspace spanned by $|0\rangle$ and $|1\rangle$ differently than the subspace spanned by $|2\rangle$ and $|3\rangle$. Indeed, we have assumed that
the initial state is a superposition of states $|0\rangle$ and  $|1\rangle$.   If the initial state is totally unknown, then it is best to construct control operators that are symmetric with respect to the two subspaces.  Upon completion of our studies of the two nesting schemes discussed in Sec. III, we find that the following three symmetry-adapted control operators can form another nesting scheme for three-layer UDD, i.e.,
\begin{eqnarray}
Z_{1}&\equiv & |0\rangle\langle 0|+|1\rangle\langle 1|-|2\rangle\langle2|-|3\rangle\langle 3|=X_{0,1} \nonumber\\
Z_{2}&\equiv& |0\rangle\langle 0|-|1\rangle\langle 1|+ |2\rangle\langle 2|-|3\rangle\langle 3| \nonumber \\
Z_{3}&\equiv & |0\rangle\langle 1|+|1\rangle\langle 0|+ |2\rangle\langle 3|+|3\rangle\langle 2|.
\end{eqnarray}
That is, for each UDD layer, the three key requirements of UDD outlined in Sec. II are satisfied. Dramatically, the form of the three control operators  $Z_{1}$, $Z_{2}$, and $Z_{3}$ is explicitly symmetric with respect to an exchange between the $(|0\rangle, |1\rangle)$ subspace and the $(|2\rangle, |3\rangle)$ subspace. So what happens to the first subspace also applies to the second subspace.
Their physical meaning is also clear: $Z_{1}$ locks the total population within each subspace spanned by $|0\rangle$ and $|1\rangle$ or
by $|2\rangle$ and $|3\rangle$, $Z_{2}$ locks the individual populations on each state, and $Z_3$ finally
suppresses the pure dephasing within each of the two subspaces of dimensional two.  Therefore, this symmetry-adapted three-layer UDD scheme should have more efficiency in reducing the algebra layer-by-layer. For the ordering of $Z_{3}-Z_{2}-Z_{1}$, the associated algebra reduction route is found to be:
\vspace{0.3cm}
\begin{center}
\framebox{$Y_i$, $i=1,2, \cdots, 16$} \\
\vspace{0.2cm}
$\Downarrow${$Z_{1}$, \text{UDD-1}}\\
\vspace{0.2cm}
\framebox{$Y_i$, $i=1,2, \cdots, 6$; $Y_{15}$, $Y_{16}$} \\
\vspace{0.2cm}
$\Downarrow${$Z_2$,\text{UDD-2}}\\
\vspace{0.2cm}
\framebox{$Y_1$, $Y_2$, $Y_3$, $Y_6$}\\
\vspace{0.2cm}
$\Downarrow${$Z_{3}$,\text{UDD-3}}\\
\vspace{0.2cm}
\framebox{$Y_1$, $Y_2$}.
\end{center}
\vspace{0.3cm}
The final effective Hamiltonian $H_{\text{eff;Z}}^{\text{UDD-3}}$ after such three UDD layers is hence a combination of only two operators: $Y_2$ and the unity operator $Y_1$ (or equivalently, $|2\rangle\langle2|+|3\rangle\langle 3|$ and $Y_1$).
Note that the dephasing between the $(|0\rangle,|1\rangle)$ subspace and the $(|2\rangle,|3\rangle)$ subspace is still not suppressed.  Perhaps even more remarkable, for these three symmetry-adapted control operators, they either commute or anti-commute, and consequently different orderings of $Z_1$, $Z_2$ and $Z_3$ can produce the same final generating algebra.

One can further rewrite $H_{\text{eff;Z}}^{\text{UDD-3}}$ in a more enlightening and symmetric form, i.e.,
\begin{eqnarray}
H_{\text{eff;Z}}^{\text{UDD-3}}&=& D_{3,1}^{Z}I \nonumber \\
& + & D_{3,2}^{Z}[|0\rangle\langle 0|+|1\rangle\langle 1|-|2\rangle\langle 2|-|3\rangle\langle 3|].
\label{HZ}
\end{eqnarray}
 This finally brings us to our last theoretical question: can we further reduce $H_{\text{eff;Z}}^{\text{UDD-3}}$ by adding one more UDD layer?  Our answer is yes in theory.   This is somewhat obvious if one introduces the fourth control operator
 \begin{eqnarray}
 Z_{4}&\equiv & |0\rangle\langle 2|+|2\rangle\langle 0|+ |1\rangle\langle 3|+|3\rangle\langle 1|
 \end{eqnarray}
 with $Z_4^2=1$.  Clearly, because $Z_4$ anti-commutes with the second operator in Eq.~(\ref{HZ}) and commutes with the identity operator, the fourth UDD layer based on $Z_4$ will finally yield an effective Hamiltonian as a certain bath operator multiplied by a unity-operator in the four-dimensional Hilbert space for a two-qubit system!  So it is theoretically possible to protect a totally unknown two-qubit state against most general system-environment coupling, using about $N^4$ control pulses in total.   Further, since this four-layer scheme is intended to lock any two-qubit state, one may now arbitrarily choose the four orthogonal basis states in order to simplify the four control operators (this is not allowed in three-layer UDD because a two-dimensional subspace is chosen beforehand).  In particular, if we consider a new set of basis states different than above, e.g., $|0\rangle=|\uparrow\uparrow\rangle$,
 $|1\rangle=|\uparrow\downarrow\rangle$, $|2\rangle=|\downarrow\uparrow\rangle$, and $|3\rangle=|\downarrow\downarrow\rangle$, then one obtains
 $Z_1=\sigma_z^1$,  $Z_2=\sigma_z^2$, $Z_3=\sigma_x^2$, and $Z_4=\sigma_x^1$, which are only local control operators in this new representation.  Such a four-layer solution is also numerically checked.

\section{Conclusion}
With both theoretical analysis and numerical study, we have shown how nested three-layer UDD can protect unknown two-qubit entangled states as a superposition of two known basis states,  to the $N$th order with about $N^3$ control pulses, without assuming a specific form of system-environment coupling.  This is of much interest to
current theoretical investigations of entanglement protection.
As a remarkable side result, it is found that the ordering of the three UDD layers can be a crucial factor.  Numerical results support our theoretical considerations.

Given the theoretical feasibility of extending UDD beyond single-qubit systems, decoherence control via nested UDD should be of experimental interest as well. Though a rigorous mathematical foundation for nested UDD is still under development \cite{Liureview,liu-multi-qubit}, the success of three-layer UDD demonstrated here in two-qubit systems further strengthens the view that nested UDD can be a good strategy for decoherence suppression.  Our approach can also be extended to protect an unknown superposition of two known basis states in an arbitrary multi-level system.

As a final extension, in Sec. IV we also discussed how a totally unknown two-qubit entangle state can be protected by applying four layers of UDD sequences that involve local operations only. Applying $N^4$ pulses can be highly demanding in experiments, but the existence of such a theoretical solution should offer a useful reference point for future studies of entanglement protection. 
 On the other hand, it becomes interesting to compare our main topic of this work, i.e., three-layer UDD, with the ultimate four-layer solution.  The four-layer solution can be realized by local control operators but the required number of pulses may present experimental difficulties. By treating a particular class of two-qubit states, three-layer UDD can achieve entanglement protection of unknown states using much less control pulses, with the price that rapid nonlocal operations are required.

\section{Acknowledgments}
We thank the Faculty of Science, National University of Singapore,  for the ``Special Program in Science" that helped to launch this research project.
J.G. is supported by the NUS ``YIA" (Grant No.
R-144-000-195-101).

\end{document}